# Many-Objective Search-Based Coverage-Guided Automatic Test Generation for Deep Neural Networks

Dongcheng Li[1], W. Eric Wong[2,*], Hu Liu[3], and Man Zhao[3]

[1]Department of Computer Science, California State Polytechnic University - Humboldt, Arcata, USA
[2]Department of Computer Science, University of Texas at Dallas, Richardson, USA
[3]School of Computer Science, China University of Geosciences, Wuhan, China
*corresponding author

*Abstract*— To ensure the reliability of DNN systems and address the test generation problem for neural networks, this paper proposes a fuzzing test generation technique based on many-objective optimization algorithms. Traditional fuzz testing employs random search, leading to lower testing efficiency and tends to generate numerous invalid test cases. By utilizing many-objective optimization techniques, effective test cases can be generated. To achieve high test coverage, this paper proposes several improvement strategies. The frequency-based fuzz sampling strategy assigns priorities based on the frequency of selection of initial data, avoiding the repetitive selection of the same data and enhancing the quality of initial data better than random sampling strategies. To address the issue that global search may yield test not satisfying semantic constraints, a local search strategy based on the Monte Carlo tree search is proposed to enhance the algorithm's local search capabilities. Furthermore, we improve the diversity of the population and the algorithm's global search capability by updating SPEA2's external archive based on a decomposition-based archiving strategy. To validate the effectiveness of the proposed approach, experiments were conducted on several public datasets and various neural network models. The results reveal that, compared to random and clustering-based sampling, the frequency-based fuzz sampling strategy provides a greater improvement in coverage rate in the later stages of iterations. On complex networks like VGG16, the improved SPEA2 algorithm increased the coverage rate by about 12% across several coverage metrics, and by approximately 40% on LeNet series networks. The experimental results also indicates that the newly generated test cases not only exhibit higher coverage rates but also generate adversarial samples that reveal model errors.

*Keywords-deep neural network; test generation; many-objective optimization; neuron coverage*

## 1. INTRODUCTION

Recent studies have shown that despite achieving high test accuracies, neural network models often exhibit unexpected or erroneous behavior in extreme conditions due to adversarial attacks, biased training data, overfitting, or underfitting. In safety-critical domains, such errors can lead to catastrophic consequences. The importance of quality and safety assurance in Deep Learning (DL) is gaining attention. Systems based on Deep Neural Networks (DNNs) must exhibit a high degree of trustworthiness before their public deployment [1].

Like traditional software, systems based on DNNs require systematic testing before deployment [2][3]. Early-stage systematic testing of DNN models to identify potential defects and vulnerabilities is crucial. While testing is a mature and common technique for traditional software, serving as a primary practice for analyzing and assessing software system quality [4][5], DL systems differ significantly in principle and structure from traditional software. The logic in traditional software is written by developers and strictly follows the source code structure. In contrast, DL models, though also defined by developers, are the result of connections between neurons and extensive dataset training. Their performance varies with changes in the network model and training set. Additionally, systems containing multiple Machine Learning (ML) models influence each other's training and tuning [6]. Hence, inspired by the traditional software testing methods, new testing methods and criteria for generating tests suitable for DNNs are required.

Common practices in existing DL testing involve collecting as much real data as possible for manual labeling or generating a large volume of simulated data to test the accuracy of neural network models. However, these methods cover only a fraction of real scenarios and do not consider the internal structure of neural network models. Moreover, manual labeling is time-consuming and labor-intensive, and using only manually labeled tests hardly builds robust DNN systems. Considering these reasons, designing an automated method for generating DL tests and improving neural network coverage has become a mainstream research direction.

Current methods for generating test inputs for neural networks mainly fall into two categories: adversarial-based and coverage-based test input generation. Fuzz testing, one of the Coverage-based methods, has become one of the main methods for generating test inputs, which creates new test inputs according to the system's input rules or/and mutation operators from existing data. Essentially, DNN test case[1] generation is an optimization problem, where a series of mutation strategies are utilized and selected to maximize model coverage. In addition, many-objective optimization algorithms have been widely used in solving various optimization problems with good effects [7].

---

[1] In this paper, we use the terms "test cases" and "test inputs" interchangeably. We also use "errors" and "defects" Interchangeably.

This paper focuses on the problem of test inputs generation for deep neural networks, aiming to generate test inputs with higher coverage, discovering defects in the neural network early in the training process, and enhancing the reliability of neural networks. The main contributions of this paper are as follows:

1) Traditional input selectors often use random selection methods for data sampling, but as time progresses, it becomes difficult to increase coverage in the later stages of testing. Therefore, this paper proposes a frequency-based fuzz data sampling strategy. That is, the probability of subsequent selections of data is set based on the number of times the data has been selected, improving the quality of initial seeds.

2) Random fuzzing methods have a vast search space, low efficiency, and significant time and space costs, making it difficult to obtain optimal solutions. A single objective heuristic search can effectively reduce the complexity of the problem and find the combinations with the maximum coverage, but they often only consider a particular state of a neuron and hard to deal with conflicts between different coverage criteria. Therefore, this paper proposed a DNN tests generation technique based on many-objective optimization method, aiming to maximize DNN coverage with various neuron coverage criteria. In addition, to address the deficiencies of many-objective optimization algorithms, improvements in local search strategies based on Monte Carlo tree search (MCTS) and decomposition-based archiving strategies are also designed and implemented.

The remainder of the paper is organized as follows. Section 2 presents related studies; Section 3 introduces the DNN test case generation as a many-objective optimization problem and describes the proposed methods; and Section 4 describes the experimental setup and analyzes the results. Finally, Section 5 presents the conclusions.

## 2. RELATED STUDIES

As deep learning technology becomes widely applied in everyday life, the reliability of neural networks has gradually gained attention. Researchers have begun proposing new tests generation techniques and coverage criteria for DNNs.

### 2.1. DNN Testing Metrics

Inspired by traditional software testing concepts, researchers believe that neural network defects can be effectively detected through specific test inputs that cover program logics. Thus, researchers have started to analyze neuron activation states, and considering neuron coverage as an important evaluation metric for DNN testing. Pei et al. [8] first introduced the concept of neuron coverage and proposed a white-box testing framework, DeepXlore, utilizing differential testing [9] to assess the quality of new test cases. Ma et al. [10] expanded on Pei's concept of neuron coverage, designing multiple criteria like K-multisection neuron coverage, strong neuron Activation coverage, neuron boundary coverage, and Top-k neuron coverage, integrating them into a multi-granularity testing framework for DL systems, DeepGauge. Inspired by traditional software, Ma et al. [11] also proposed a coverage-guided test generation framework, DeepCT, based on Combinatorial Testing and two testing criteria - t-way sparse coverage and t-way dense coverage.

Zhou et al. [12] introduced the contribution coverage criterion, differing from traditional neuron coverage, describing the combination of a neuron's output and its emitted connection weights. He further proposed a contribution coverage-guided fuzz testing framework, DeepCon-Gen, treating test case generation as an optimization problem solved by gradient-based optimization. Inspired by traditional software MC/DC coverage standards, Sun et al. [13] introduced SS (Symbol-Symbol) coverage to detect symbol changes in DNN features, along with VS (Value-Symbol), SV (Symbol-Value), VV (Value-Value) coverage. Inspired by path-guided testing, Wang et al. [14] proposed a series of path-driven testing standards, DeepPath, for comprehensive coverage computation in deep neural networks. Viewing deep neural networks as weighted directed graphs of nodes and edges, they introduced l-SAP, l-OAP, and l-FSP as testing criteria.

Besides CNN research, Du et al. [15] proposed an automated testing framework for Recurrent Neural Networks (RNNs), DeepCruiser, modeling RNNs as abstract state transition systems and defining a set of testing coverage criteria specifically for stateful DL systems. Kim et al. [16] believed existing neuron coverage and other criteria were not fine-grained enough to capture subtle behaviors of DL systems, thus they proposed a testing framework SADL for DL systems, which introduced surprise adequacy as a testing criterion.

### 2.2. Fuzz-based Test Generation Techniques for DNN

Fuzzing is a widely used technique for exposing software defects, balancing software testing efficiency and effectiveness through code coverage feedback. Guo et al. [17] proposed the first differential fuzz testing framework for DL, DLFuzz, creating minor perturbations to original images to change test inputs and activate more neurons, covering more system logic. They set minor perturbation constraints to ensure consistent prediction results before and after mutation and used differential testing to automatically identify erroneous program behaviors. Zhang et al. [18] used differential testing to detect DNN system behaviors for identical inputs, employing neuron coverage as a measurement criterion. To ensure generated test inputs remained recognizable, they proposed a semantic transformation constraint strategy, converting behavior divergence and neuron coverage into a multi-objective problem solved using the NSGAII algorithm. Odena et al. [19] introduced the concept of coverage-guided fuzzing into neural network testing, created an open-source deep learning neural network testing framework, TensorFuzz. It obtains input-output from TensorFlow's computational graph, generates new test inputs by adding noise to test data, and uses approximate nearest neighbor algorithms to describe neuron coverage.

Xie et al. [20] believed traditional fuzz testing frameworks oversimplified system input mutations, failing to detect complex defects in DNN systems. They designed and implemented a new mutation strategy to generate test inputs, integrating it into the proposed coverage-guided fuzz testing framework, DeepHunter, for detecting latent defects in general DNNs. It uses multiple scalable coverage criteria as feedback to guide test generation. Demir et al. [21] transformed the neural network test inputs generation problem into finding a series of mutation operations that maximize coverage, using Monte Carlo tree search to drive coverage-guided testing, and proposed a coverage-guided fuzz testing framework for DNN structural testing, DeepSmartFuzzer. Zhang [22] proposed a coverage-guided adversarial tests generation framework for deep learning systems, CAGFuzz, training an Adversarial Example Generator (AEG) on a general dataset, considering only data features to avoid poor generalization issues, extracting generative features of original and adversarial data, and constraining adversarial data through cosine similarity to preserve semantic information.

Tian et al. [23] proposed an automated synthetic test case generation technique for testing autonomous driving systems, using neuron coverage as a metric. Using various image transformation techniques, they simulated different real driving conditions (blurring, rain, fog, etc.) to generate synthetic test cases that increased neuron coverage. Gao et al. [24] proposed a mutation-based fuzz testing method to generate new test cases to enhance DNN system robustness. The method tackled DNN system generalization data augmentation issues through guided test generation techniques, turning the test case generation problem into an optimization problem, and conducting genetic searches across the variable space of each training input data, selecting data that performed worst on DNNs for enhancement.

Braiek et al. [25] proposed a search-based fuzz testing framework for DNN models, DeepEvolution, relying on population-based meta-heuristic methods to explore the search space of semantic-preserving transformations, using a coverage-based fitness function to guide the exploration process. Kang et al. [26] introduced a new search method, not searching the entire image space but a reasonable space similar to the formal training distribution, using SINVAD (Search-based Input Space Navigation using Variational Autoencoders) to search the image space and generate test cases, and proposed a fuzz testing generation technique based on genetic algorithms. Yi et al. [27] designed a neural network fuzzing framework based on genetic algorithms, selecting neuron coverage as a coverage criteria. They provided a new data sampling method for selecting algorithm inputs, ensuring higher coverage from selected data. Pour et al. [28] proposed a search-based fuzz testing framework for DNNs involved in source code embedding and code search. They used popular source code refactoring tools to generate semantically equivalent variants, guiding test searches with DNN mutation testing.

Guo et al. [29] introduced Audee, a new method for testing DL frameworks and localizing defects. Audee employed a search-based approach, exploring combinations of model structure, parameters, weights, and inputs, implementing three different mutation strategies to generate diverse test cases. Huang et al. [30] conducted an in-depth study on a class of Recurrent Neural Networks, Long Short-Term Memory networks, and developed a coverage-guided testing tool, TestRNN. Addressing the issue of existing metrics failing to cover the internal structure of RNNs, they added temporal relationship indicators including step length and bounded length, using random mutation and goal-oriented mutation based on genetic algorithms to generate new test cases. Hu et al. [31] proposed a test case generation method based on differential evolution (DE), with evolutionary algorithms offering better global optimization capabilities than gradient optimization. This method used DNN's prediction loss and selected coverage criteria as fitness functions, performing minor perturbations between different channels of an image to construct test cases.

## 3. MANY-OBJECTIVE SEARCH-BASED DNN TEST GENERATION

The problem of generating test cases for neural networks fundamentally belongs to the domain of optimization problems. For images, it can be divided into multiple local areas, each of which can undergo various mutation operations. The aim of this paper is to optimize the following objectives by adding perturbations to original test data to generate new test inputs:

$$\begin{cases} Maximize\ KMNC(T, input) \\ Maximize\ NBC(T, input) \\ Maximize\ SNAC(T, input) \end{cases} \quad (1)$$

Here, *input* represents the original test data, and $T$ denotes the target DNN system. KMNC, NBC, and SNAC respectively stand for k-multisection neuron coverage, boundary neuron coverage, and strong neuron activation coverage [10]. Due to the different ranges of neuron activation values corresponding to various coverage criteria, conflicts exist between these criteria. For example, as the coverage of a neuron's primary functional area increases, the coverage of its boundary area decreases. Hence, the problem of generating test cases for neural networks can be transformed into a many-objective optimization problem.

Due to the significant differences in mutation operations among images, audio, and natural language, this paper mainly focuses on convolutional neural networks (which is a type of deep neural network) and primarily targets image datasets. The chromosome encoding is shown in Figure 1. Each chromosome consists of three parts: the first part is a vector of length n, representing different areas of the image, encoded in binary, where 1 indicates the selection of that area for mutation. Considering that global mutation methods can significantly impact pixel values and may alter the image's semantics, local mutation methods are used to ensure diversity in the images. The value of n needs to be determined based on the specific size of the images in different datasets.

The second part of the chromosome represents mutation features for each area. A major challenge in designing effective mutation strategies for coverage-guided fuzzing (CGF) is balancing the mutability of the strategy with image semantic constraints. If the mutation strategy changes the image pixels too drastically, the semantics of the image may change or even become unrecognizable to the neural network; if the change is too slight, the newly generated test cases will not differ meaningfully from the original ones. Traditional image mutation strategies mainly fall into two categories [32]: 1) Pixel value transformations: image contrast, blur, brightness, noise, and 2) Affine transformations: image translation, scaling, shearing, rotation.

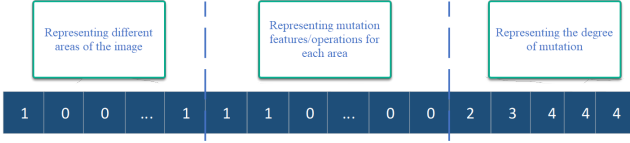

Figure 1. Chromosome Encoding

In essence, pixel value transformations change the pixel values of an image, while affine transformations move the pixels. These strategies have been proven effective through related research [32]. Since operations based on affine transformations may render images unrecognizable, this paper mainly uses pixel value transformations, including image contrast, blur, and brightness, as mutation strategies.

Given an image $s$, $t$ is a class of mutation strategies, $s'$ is the image after one local transformation, i.e., $s \xrightarrow{t} s'$, and local mutations based on pixel values generally increase variability through multiple applications, i.e., $s \xrightarrow{t_0, t_1, \dots t_n} s'$. To prevent the image from becoming unrecognizable after multiple mutations, the following constraint is defined [21]:

$$f(s,s') \begin{cases} L_0(s,s') \leq 255 & if (L_0(s,s') < \propto \times \ size(s)) \\ L_\infty(s,s') < \beta \times 255 & otherwise \end{cases} \quad (2)$$

Where $0 < \propto, \beta < 1$, $L_0(s,s')$ represents the maximum pixel change between images $s$ and $s'$, $L_\infty(s,s')$ is the maximum pixel transformation value, and $size(s)$ represents the total number of pixels in the image. That is, when the number of changed pixels in an image is very small (less than $\propto \times size(s)$), it is considered not to affect semantics, ensuring pixel values remain valid (less than 255). When the number of pixel changes is too large, the maximum pixel change should be limited (less than $\beta \times 255$).

The third part of the chromosome represents the degree of mutation for each mutation feature. For each mutation operation, multiple mutation levels are set. For example, a chromosome of $(1, 0, 1, 2, 1, 0, 2, 3, 4)$ indicates that for the first area of the image, the second level of image brightness transformation is applied, for the third area, the fourth level of image contrast transformation is applied, and for the second area, no change is made. The specific mutation process is shown in Algorithm 1.

| **Algorithm 1**: Mutation Process |
| --- |
| Input: s: Initial Data Set，K: Total Number of Samples |
| Output: T: Generated Test Cases |
| $(s_0, s_0') = Info(s)$ //Loading the Data Set |
| T = [] |
| for j in range(0, k): |
|     action1: Randomly Acquiring a Region of the Image |
|     action2: Random Selection of a Mutation Strategy |
|     $s_0' = apply(action1, action2, s_0)$ // Applying Mutation to the Region of Image $s_0$ |
|     while isSatisfied($f(s,s')$) and reward > 0 then //Satisfying Semantic Constraints and Increasing Coverage |

| **Algorithm 1**: Mutation Process |
| --- |
|     action1, action2 = select (area, categories) //Selecting a New Region and Mutation Strategy |
|     $s_0' = apply(action1, action2, s_0')$ //Applying the Mutation to a Local Region of the Image |
|     If $s_0'$ is the not same with $s_0$ then: //If the Newly Generated Test Case Differs from the Original Image |
|         T = T ∪ $s_0'$ |
| return T |

## 3.1. Fuzz Testing Framework for Neural Networks Based on Many-Objective Search

Fuzz testing and adversarial testing are currently mainstream DNN test generation methods, primarily targeting convolutional neural networks, with objective functions often being neuron coverage and its variants such as k-multisection neuron coverage, etc. However, most existing frameworks adopt random or gradient-based methods, or single-objective heuristics for solution searching. Random methods have a large search space, gradient-based optimization relies too heavily on the quality of gradient functions, and single-objective heuristic searches only consider one type of neuron state. Addressing these shortcomings in existing frameworks, this paper proposes a fuzz testing framework for neural network test input generation based on many-objective optimization algorithms, using heuristic searches to effectively explore the mutation space and employing various coverage criteria as guiding feedback to further enhance neuron coverage.

The overall framework, as shown in Figure 2, mainly consists of four parts: neural network model and dataset preprocessing module, data selection module, many-objective optimization algorithm module, and adversarial analysis module.

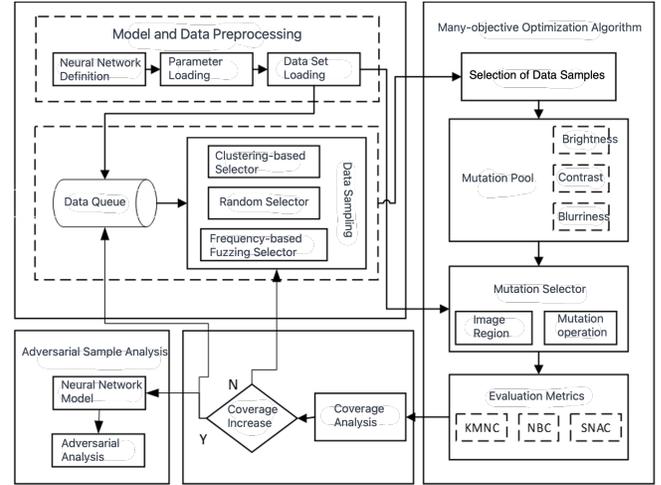

Figure 2. Framework Design

Neural Network Model and Dataset Preprocessing Module: This mainly includes the definition of the neural network model under test, loading of the original dataset, and configuration of algorithm parameters and evaluation metrics, etc.

Data Selection Module: Considering the large size of deep neural network datasets, it is necessary to fetch images in batches from the dataset for mutation analysis. Traditional testing frameworks generally obtain mutation input batch batch randomly, which is inefficient and contributes little to coverage improvement in later stages of sampling. This paper proposes a frequency-based fuzzy sampling strategy to select mutation inputs, ensuring that the probability of previously mutated images being selected again decreases over time, thereby effectively enhancing test case coverage in later iterations.

Algorithm Module: Local mutation methods do not guarantee diversity in image mutations, possibly resulting in newly generated test cases that are not significantly different from the originals. Therefore, a series of mutation strategies $M$ are combined to generate new mutated inputs batch $batch'$, and random mutations, lacking effective guidance, perform poorly in search. This paper uses k-multisection neuron coverage, strong neuron activation coverage, boundary neuron coverage, as guiding generation criteria, employing image blur, brightness, and contrast as mutation strategies. Each strategy is set with multiple levels based on the degree of mutation, and due to the numerous combinations and a large search space, a many-objective optimization algorithm is used to find best batch $batch_{best}$, the combination that maximizes coverage in a short amount of time. By assessing the coverage increase of the mutated images, if they effectively enhance coverage, they are re-added to the original dataset.

This paper uses the SPEA2 algorithm, a method based on Pareto dominance, for solving the DNN test generation problem. Convergence and diversity are two important performance requirements in the algorithm. Algorithms based on Pareto dominance improve the algorithm's convergence performance in many-objective problems. SPEA2 introduces the concept of an external archive and uses a density distribution method to enhance the algorithm's diversity, making the algorithm's solutions more closely approximate the real PF (Pareto Front) in solving neural network test case generation problems. The algorithm has fewer parameters than indicator-based algorithms, which consume a lot of time calculating various indicators, hence offering higher efficiency in neural network test case generation.

Adversarial Analysis Module: To further verify the effectiveness of the test cases generated through the framework, i.e., whether the new test cases not only achieve high coverage but also effectively expose errors or defects in the DNN system. It involves adversarial analysis of the newly generated test samples, inputting them into the target DNN system for testing, and counting the number of generated adversarial samples.

## 3.2. DNN Test Generation using Frequency-based Fuzzing Sampling

Mutation-based fuzzer randomly mutate existing seeds to generate new ones, but random mutation methods produce many useless seeds and can only discover a small portion of vulnerabilities in complex programs. Generation-based fuzzer have high structural requirements for inputs and limited generation rules that restrict the search space of input programs.

Using software coverage information to guide fuzz searches has improved testing performance, allowing for the discovery of more defects in programs. However, adding coverage information alone is not sufficient to cover all program paths and the quality of the initial seeds have a great impact on the performance of fuzz testing [33]. The main factors affecting fuzzers' effectiveness are as follows:

1) Quality of initial seeds: The quality of initial seeds is significant for subsequent mutations, and it is difficult to construct effective seeds based on random sampling.
2) Seed selection strategy: Effective data sampling strategies can lead to higher coverage seeds being selected, avoiding the repeated selection of useless seeds and impacting the efficiency and performance of test case generation.
3) Mutation strategy: Coverage information can guide test cases to search towards higher coverage, but random mutation strategies often produce useless test cases and hardly increase coverage effectively.

Therefore, this paper designs a frequency-based fuzzing sampling strategy based on the number of times a seed is selected. To achieve high neuron coverage in CGF, it is necessary to prioritize the seeds to ensure that seeds that increase coverage are selected with a higher probability. Therefore, the probability of selection should meet the following requirements:

1) Probability values are kept between (0, 1).
2) The more times a seed is selected, the lower the probability of its being selected again.

For this purpose, the following function is designed:

$$P_1(x) = \frac{1}{1+e^{ax+b}} \ a > 0, x \in [0, +\infty) \quad (3)$$

Where x is the number of times a seed is selected, and $a$ and $b$ are set as hyperparameters.

Additionally, in the middle and later stages of testing, the neuron coverage rate generated by test inputs becomes lower. These lower neuron coverage rates are more valuable at this time, as test cases that generate new coverage rates are more likely to discover defects in programs. Therefore, another seed selection function is designed:

$$P_2(NCov_{new}, Circle, Total) = {}^{1+\frac{Circle^4}{Total^3}}\sqrt{NCov_{new}} \quad (4)$$

Where $NCov_{new}$ is the increased coverage rate in the last iteration, $Circle$ is the current iteration number, and $Total$ is the target iteration number. When $Circle$ is small, $P_2$ approximates a linear function, making it easier to select seeds that increase coverage. When $Circle$ is large, $P_2$ better reflects the relationship between smaller coverage rate increases and seed selection probabilities.

Finally, the following formula can be obtained:

$$P = \left(1 - \frac{Circle}{Total}\right) \times P_1 + \frac{Circle}{Total} \times P_2 \quad (5)$$

The pseudo-code for the DNN test case generation process using frequency-based fuzzing strategy is shown in Algorithm 2.

| **Algorithm 2**: DNN Test Case Generation Process using Frequency-based Fuzzing Sampling Strategy |
|---|
| Input: I: Initial Seeds, N: Target Neural Network |
| Output: T: A set of new generated tests |
| Const: K: A configurable total number of seed mutation |
| T = [] |
| Q = [I] //Original Data Set |

| **Algorithm 2**: DNN Test Case Generation Process using Frequency-based Fuzzing Sampling Strategy |
|---|
| SelectNext = selects a seed based on the number of times it has been fuzzed |
|   While s = SelectNext(Q): |
|       T = MetamorphicMutate(s, K) //Select Mutation Strategy from the Mutation Pool |
|       Cov, result = run(DNN, T) //Input the New Test Case into the Target Neural Network for Prediction |
|       For $s' \in T$ do |
|         If isCoverageAdd(cov) then //If Coverage Increases, Add to Set T |
|           T = T ∪ $s'$ |
|         If isAdversarialExamples($s'$, $result$) //Determine if It Is an Adversarial Sample |
|           Q = Q ∪ $s'$ |
|   Return T |
|   End |

### 3.3. Local Search using MCTS for SPEA2 based DNN Test Generation

The basic idea of local search is to start with a random or heuristically generated candidate solution for a given problem instance, search its neighborhood, and iteratively improve this candidate solution through minor modifications.

This paper proposes a DNN test case generation method based on SPEA2. Each individual represents a set of test cases, and the chromosome encoding is divided into 3 parts (as discussed in previous section), finding an optimal set of mutations for the local areas of the image to maximize neuron coverage. However, during the population iteration process, selecting multiple areas for mutation at once causes chromosomes to evolve into solutions that do not meet image semantic constraints, resulting in too many invalid solutions, rapid convergence of the population, and insufficient diversity. To address these issues, a local search algorithm based on MCTS is proposed. Each time, only one local area of the input image is mutated, and whether to select another area for mutation is decided based on whether new coverage is generated. This method avoids many individuals failing to meet semantic constraints, increases the diversity of the population, enhances the algorithm's local search capability, and generates test cases with higher coverage.

MCTS is a heuristic algorithm used mainly in board game software and decision-making. It uses a search tree to represent the game, where each node of the tree represents a specific state of the game. When performing a certain operation, it transitions from one game state to its child node. The MCTS algorithm is used to select the most advantageous action in any state of the game, with the ultimate goal of finding the optimal path (sequence of actions) to win the game.

The coverage-guided fuzzer driven by MCTS consists of an input selector, coverage criteria, input mutator, and mutation selector [21]. For each iteration, the input selector selects a batch of inputs $I$ from the original dataset according to a specific data sampling strategy. Then, the mutation selector determines which mutation combination $(r', m')$ to apply to the input, where $r'$ represents a certain local area of the input image, and $m'$ represents one of the mutation strategies. The input $I$ and the mutation combination $(r', m')$ are added to the input mutator to form a new input, i.e., $I \xrightarrow{(r',m')} I'$. The new input $I'$ is put into the neural network model for prediction, and the coverage rate is calculated according to the coverage criteria formula. If the coverage rate is improved, it continues to be added to the mutation selector, and a new mutation $(r'', m'')$ is selected to form a new input, i.e., $I' \xrightarrow{(r'',m'')} I''$, until the constraint conditions or the algorithm's termination conditions are not met. If the final input increases coverage, it is added to the new test set, and then the input selector is used to select a new batch of data from the original dataset for mutation processing.

The entire mutation process can be abstracted into a game-like problem and solved using the MCTS method. The game tree nodes correspond to image areas and mutation strategies, i.e., player 1 selects the local area of the image to be mutated, and player 2 selects the mutation operation to be performed in that area. Then, continuous selection of image area and mutation operation is performed to calculate the coverage criteria, updating the generated input with the highest coverage, and re-adding it into the test set after the game ends. The tuple of actions taken by players 1 and 2 is called a complete behavior $(r, m)$, with odd layers of the game tree being the local areas of the image and even layers being mutation operations such as brightness, blur, contrast, etc., each operation set with multiple levels.

The algorithmic process of SPEA2 using MCTS local search is shown in Figure 3. The local search based on MCTS is used for neighborhood search when global mutations cause chromosomes to fail to meet image semantic constraints. Each time, only one local area of the image and one mutant are searched, and a new area and mutation operation are only selected if new coverage is generated. As the search process of MCTS is time-consuming, and its search effect will deteriorate in the later stages of test case generation, it is necessary to dynamically control the proportion of chromosomes undergoing local search. A threshold can be used for control, and a linear transformation operation is adopted. The greater the number of population iterations, the greater the threshold. At this time, the global search capability of the algorithm is used to select mutation operations and areas to generate higher coverage.

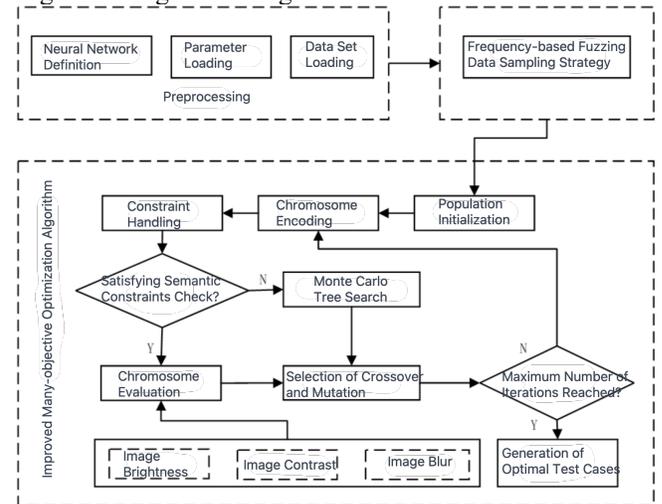

Figure 3. SPEA2's Local Search using MCTS

The pseudo-code for the DNN test case generation process based on SPEA2 using MCTS is shown in Algorithm 3:

**Algorithm 3**: DNN Test Case Generation Process Based on SPEA2 using MCTS

Input: List population , List Archive, Int T(Number of Iterations), game, time(Test Case Generation Time)

Output: T: A set of new generated tests

T = []

While t < time:

    I = SelectNext(dataset) //Selecting a Batch of Experimental Samples from the Dataset

    Population, archive = Initialize() //Initializing the Population and External Archive

    for i in range(T):

        population = append(archive)

        calfitness(population) //Calculating Population Fitness

        archive = selection(population)

        coverage_best, $I^{best}$ = updateBest(archive) //Updating Optimal Chromosomes and Fitness Values

        mp = selectionTournoi(archive)

        enfant = SimulatedBinaryCrossover(mp)

        pop = PolynomiaMutation(enfant)

        for j in len(pop):

            if pop[j] isn't meet constraints and random > C:

                root = MCTS_Node(game)

                $(r',m')$ = run(MCTS)

                pop[j] = Individual$(r',m')$

            else:

                pop[j] = []

        population = evaluate(pop)

## 3.4. Improve DNN Test Generation with Decomposition-based Archiving

The SPEA2 algorithm employs an external archive to store the superior individuals of a population, ensuring its diversity. However, the calculation of non-dominated solutions requires considering the dominance relations of both the population and the external archive, leading to an exponential increase in computational cost with the growth of the archive size. To better balance the diversity and convergence of the population, this paper proposes a decomposition-based archiving method to enhance the global search capability of the SPEA2 algorithm for DNN test case generation.

The decomposition-based archiving method uses a set of uniform weight vectors to divide the objective space into multiple subspaces. During the population's evolution, the best-found solutions are saved in the corresponding subspaces. The size of the archive equals that of the subspaces, and normalized distances are used to determine the subspaces to which solutions belong, facilitating comparisons of dominance relations without computing all solutions, thereby significantly saving computational cost.

Besides dividing weight vectors, updating the external archive is also important. This paper mainly uses the normalized distance method to update the archive, involving two steps: first, determining the weight vector associated with each solution, where solutions with the same weight vector are considered to be in the same subspace; second, pruning the external archive by deleting redundant solutions from each subspace. The main process of the algorithm is as shown in Figure 4.

The preprocessing and sampling strategies from the previous section remain the same and are not reiterated here. Initially, the algorithm parameters, including the number of target vectors, are initialized, and the initial population and external population are randomized. Next, to design the weight vectors, involves dividing them into a three-dimensional hyperplane, followed by constraint processing, including pixel value constraints and image semantic constraints, and the calculation of fitness values, such as k-multisection neuron coverage, boundary neuron coverage, etc. Unlike traditional SPEA2, this method does not compare the dominance relations of the population and external archive simultaneously. Instead, it calculates the normalized distance between the population and the weight vectors to determine the subspace of the solutions. Then, redundant non-dominated solutions are deleted from each subspace, saving the best solutions in the archive. Binary tournament selection, crossover, and mutation operations are then used to generate the next generation of the population until the termination conditions are met to output the Pareto optimal solutions.

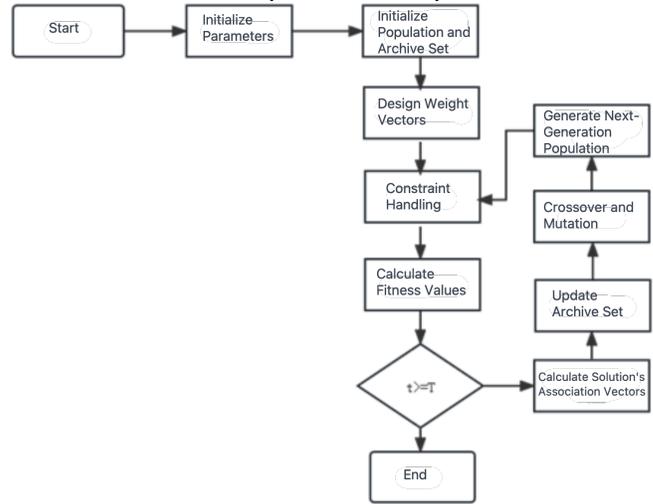

Figure 4. The flowchart of SPEA2 with decomposition-based Archiving Strategy

The pseudocode of SPEA2 with decomposition-based archiving strategy is shown in Algorithm 4.

**Algorithm 4**: SPEA2 with Decomposition-based Archiving Strategy

Input: population, N(Population Size), $\lambda$ (Weight Vector), T(Maximum Number of Iterations), $Z^*$(Reference Point)

Output: Non-dominated Optimal Solution Set

Population = Initialize();

Pop = $\{P_1, P_2, \ldots \ldots P_k\}$ //Division According to Weight Vector

while I < T

    population = population ∪ Pop

    calFitness(population) //Calculation of Population Fitness Values

    if reference point $Z^*$ is update:

        re-calcualte the $\lambda$ //Re-updating the Weight Vector

    calculate the distance between the i-th solution in population with Pop

    find the all non-dominated solutions in Pop

    update the reference point $Z^*$: $Z^*$ = min $\{f_i(X)\}$

    nextPop = update the Pop (Eliminate Redundant Solutions)

    Select Individuals from nextPop and population Using Binary Tournament Method

    Perform Crossover and Mutation to Generate the Next Generation Population

    End

## 4. EXPERIMENTAL DESIGN AND RESULT ANALYSIS

To validate the effectiveness of the DNN test generation method proposed in this paper, as well as various improvement strategies for algorithms and frameworks, experiments and analyses were conducted on several neural networks such as LeNet1, LeNet4, LeNet5, and on multiple public datasets including MNIST, CIFAR10, among others. Comparisons with popular DNN testing frameworks were also made. Finally, an adversarial analysis of the newly generated test cases was conducted to verify the effectiveness of the test inputs generated by the proposed method.

### 4.1. Experimental Design

#### 4.1.1. Research Questions

The following research questions are addressed through experimental result analysis:

1) RQ 1: The performance of using a frequency-based fuzz sampling strategy compared to random and clustered searches in terms of coverage improvement under various DNN testing criteria.
2) RQ 2: The performance of the test generation technique based on the many-objective search in solving the problem of DNN test cases generation.
3) RQ 3: The effectiveness of the improved SPEA2 with MCTS-based local search in DNN test cases generation under multiple coverage criteria.
4) RQ 4: The effectiveness of the improved SPEA2 using a decomposition-based archiving strategy in DNN test cases generation under multiple coverage criteria.
5) RQ 5: The effectiveness of the test inputs generated by the proposed method to the original model.

#### 4.1.2. Experiment Setup

The experimental setup is detailed in Table 1. To ensure accuracy and universality of the test results, each experiment was conducted multiple times, and the final results were averaged.

Table 1. Experimental setup

| Testing Platform | Windows |
| --- | --- |
| Hardware | Intel Core i9-10900F@ 2.80GHz RAM: 64G GPU: RTX3070 |
| Software | Pycharm2019 Tensorflow 2.4.0 CUDA11.3 |

#### 4.1.3. Datasets and Models

In this study, neural network models LeNet1, LeNet4, LeNet5, CIFAR-CNN, VGG16, and VGG19 were selected. Datasets including MNIST, CIFAR-10, and ImageNet were chosen. Details of datasets and neural network model parameters are shown in Table 2.

#### 4.1.4. Experimental Objective Functions

Objective functions chosen for optimizing neural network test case generation included recent neuron coverage related criteria such as KMNC, SNAC, and NBC. Traditional neuron coverage, which uses a threshold to determine neuron activation, often yields results highly dependent on the chosen threshold value. To address this, multiple neuron coverage criteria were considered to make the coverage criteria more granular and universally applicable.

Table 2. Datasets and Neural Network Model Parameters

| Dataset | Description | DNN Name | Neurons | Accuracy |
| --- | --- | --- | --- | --- |
| MNIST | Hand Written Digits from 0 to 9 | LeNet1 | 52 | 98.33% |
| | | LeNet4 | 148 | 98.59% |
| | | LeNet5 | 268 | 98.96% |
| CIFAR-10 | 10 class | CIFAR-CNN | 3152 | 77.68% |
| ImageNet | 1000 class | VGG-16 | 14888 | 92.8% |
| | | VGG-19 | 16168 | 92.7% |

#### 4.1.5. Algorithm Parameter

Tables 3 and 4 show the optimal parameter configurations for the many-objective optimization algorithms used in this study, determined through comparative experiments.

Table 3. Parameter Configuration for SPEA2

| Parameter Names | Values |
| --- | --- |
| Population Size | 40 |
| External Archive Size | 40 |
| Number of Iterations | 40 |
| Mutation Probability | 0.2 |
| Coverage Criteria | KMNC、NBC、SNAC |

Table 4. Parameter Configuration for NSGAIII

| Parameter Names | Values |
| --- | --- |
| Population Size | 40 |
| Number of Iterations | 40 |
| Mutation Probability | 0.2 |
| Coverage Criteria | KMNC、NBC、SNAC |

#### 4.1.6. Experimental Design

To verify the effectiveness of the many-objective search-based method and improvement strategies for DNN test generation, SPEA2 and NSGAIII algorithms were used for comparison with other DNN test generation methods. Frameworks and strategy combinations used are shown in Table 5.

Table 5. Framework and Strategy Combinations

| Framework Name | Base Framework | Employed Strategies |
| --- | --- | --- |
| DSF-Random | DeepSmartFuzzer | Random Sampling |
| DSF-clustered | DeepSmartFuzzer | Clustering-based Sampling |
| DSF-Prob | DeepSmartFuzzer | Frequency-based Fuzzing Sampling |
| DeepSearch(MSPEA2) | DeepSmartFuzzer | SPEA2 with MCTS local search |
| DeepSearch(DSPEA2) | DeepSmartFuzer | SPEA2 with Decomposition-based Archiving |
| DeepSearch(NSGAIII) | DeepSmartFuzzer | NSGAIII |
| DeepHunter [20] | DeepHunter | Fuzzing |
| Tensorfuzz [19] | Tensorfuzz | Fuzzing |

## 4.2. Experimental Results and Analysis

### 4.2.1. The effectiveness of Frequency-Based Fuzz Sampling Strategy for DNN test generation

To validate the effectiveness of the frequency-based fuzzy sampling strategy, experiments were conducted using DeepSmartFuzz framework. Experiments were conducted with 16 samples per framework, choosing 256 images per sample. Results are shown in Tables 6 to 11.

Table 6. Multi-criteria coverage of various sampling strategies on the LeNet1 network in the MNIST dataset

| Model/Dataset | Framework | KMNC | NBC | SNAC |
|---|---|---|---|---|
| LeNet-1 MNIST | Initial | 37.79% | 6.25% | 6.25 |
| | DSF-Random | 79.86% | 45.83% | 47.91% |
| | DSF-clustered | 74.93% | 28.12% | 37.5% |
| | DSF-Prob | **86.11%** | **50.04%** | **50.13%** |

Table 7. Multi-criteria coverage of various sampling strategies on the LeNet4 network in the MNIST dataset

| Model/Dataset | Framework | KMNC | NBC | SNAC |
|---|---|---|---|---|
| LeNet-4 MNIST | Initial | 39.57% | 9.51% | 14.08% |
| | DSF-Random | 80.7% | 27.46% | 40.14% |
| | DSF-clustered | 75.71% | 24.1% | 39.43% |
| | DSF-Prob | **85.02%** | **30.64%** | **44.376%** |

Table 8. Multi-criteria coverage of various sampling strategies on the LeNet5 network in the MNIST dataset

| Model/Dataset | Framework | KMNC | NBC | SNAC |
|---|---|---|---|---|
| LeNet-5 MNIST | Initial | 36.27% | 8.78% | 15.27% |
| | DSF-Random | 77.17% | 26.9% | 41.98% |
| | DSF-clustered | 79.7% | **32.89%** | 44.99% |
| | DSF-Prob | **81.47%** | 32.26% | **50.79%** |

Table 9. Multi-criteria coverage of various sampling strategies on the CIFAR-CNN network in the CIFAR10 dataset

| Model/Dataset | Framework | KMNC | NBC | SNAC |
|---|---|---|---|---|
| CIFAR-CNN CIFAR10 | Initial | 25.42% | 4.89% | 9.14% |
| | DSF-Random | 54.42% | 15.04% | 40.3% |
| | DSF-clustered | 53.39% | 18.79% | **41.03%** |
| | DSF-Prob | **62.61%** | **19.2%** | 37.49% |

Table 10. Multi-criteria coverage of various sampling strategies on the VGG16 network in the ImageNet dataset

| Model/Dataset | Framework | KMNC | NBC | SNAC |
|---|---|---|---|---|
| VGG16 ImageNet | Initial | 24.59% | 4.51% | 6.19% |
| | DSF-Random | 46.78% | 15.48% | 20.53% |
| | DSF-clustered | 50.51% | 18.45% | 22.81% |
| | DSF-Prob | **50.85%** | **21.73%** | **24.87%** |

Table 11. Multi-criteria coverage of various sampling strategies on the VGG19 network in the ImageNet dataset

| Model/Dataset | Framework | KMNC | NBC | SNAC |
|---|---|---|---|---|
| VGG19 ImageNet | Initial | 22.27% | 5.32% | 7.64% |
| | DSF-Random | 44.51% | 15.63% | 19.32% |
| | DSF-clustered | 46.85% | 17.5% | **23.51%** |
| | DSF-Prob | **48.52%** | **22.68%** | 21.83% |

Tables 6 to 11 show the effects of various sampling strategies like DSF-Random, DSF-clustered, DSF-Prob on the coverage criteria across different neural networks and datasets. From the experimental results, we can draw the following two conclusions:

(1) The DSF-Prob strategy proposed in this paper shows a certain improvement, and in most cases, it performs better compared to random and clustering-based sampling strategies, although it underperforms clustering-based sampling in some network models. DSF-Prob primarily ensures data is uniformly selected by setting priorities, resulting in effective coverage rate improvement even in the later stages of iteration. The clustering-based strategy - DSF-clustered, which samples based on image features, shows little difference in output and thus a smaller improvement in coverage rate for simple network structures. However, as network models become more complex, its coverage rate improvement surpasses random sampling but is slightly less effective than the frequency-based fuzzing strategy - DSF-Prob. The random sampling strategy - DSF-Random, when applied to smaller datasets and relatively simpler network structures, results in a relatively larger coverage rate improvement due to the large overall sampling size. As datasets become more complex, its effectiveness decreases compared to clustering-based and frequency-based fuzzing strategies.

(2) Since the KMNC metric primarily describes the main functional area of neuron activation values, there is a significant improvement in the coverage rate. For example, in the LeNet series networks, the DSF-Prob strategy increased the coverage rate by about 50%, as seen in Table 6, SNAC and NBC, which mainly describe the boundary areas of neurons, start with a lower initial coverage rate and achieve a maximum increase of 45% in the LeNet series networks. With network models' complex increases, the improvement in coverage rate decreases, especially for the VGG series networks. As shown in Tables 10 and 11, due to the ImageNet's large training set, large number of neurons, large neuron activation thresholds, and a small sampling size relative to the entire validation set, the overall coverage rate improvement is lower than for the LeNet series networks and the CIFAR_CNN model.

In short, the DSF-Prob sampling strategy is effective in improving various neuron coverage criteria for DNNs across different datasets. As the complexity of neural network models and datasets increases, the improvement in coverage gradually decreases, but it still performs better than random and clustering-based sampling in most cases

### 4.2.2. Performance of Many-Objective Test Case Generation

To validate the effectiveness of many-objective optimization for DNN test cases generation, SPEA2, NSGAIII, and MCTS (used by the original framework, DeepSmartFuzz, as the primary search) algorithms were compared. Results of the experimental comparisons are shown in Tables 12 to 17.

Table 12. Comparison of search performance of various algorithms on the LeNet1 network in the MNIST dataset

| Model/Dataset | Strategies | KMNC | NBC | SNAC |
|---|---|---|---|---|
| LeNet-1 MNIST | Initial | 37.79% | 6.25% | 6.25% |
| | SPEA2 | 66.87% | **31.26%** | **25.01%** |
| | NSGAIII | 66.21% | 20.44% | 22.91% |
| | MCTS | **68.31%** | 28.3% | 18.76% |

Table 13. Comparison of search performance of various algorithms on the LeNet4 network in the MNIST dataset

| Model/Dataset | Strategies | KMNC | NBC | SNAC |
|---|---|---|---|---|
| LeNet-4 MNIST | Initial | 39.57% | 9.51% | 14.08% |
| | SPEA2 | **71.1%** | 21.06% | **28.94%** |
| | NSGAIII | 64.83% | **23.61%** | 22.32% |
| | MCTS | 70.89% | 21.71% | 25.29% |

Table 14. Comparison of search performance of various algorithms on the LeNet5 network in the MNIST dataset

| Model/Dataset | Strategies | KMNC | NBC | SNAC |
|---|---|---|---|---|
| LeNet-5 MNIST | Initial | 36.27% | 8.78% | 15.27% |
| | SPEA2 | **67.54%** | 17.94% | 30.53% |
| | NSGAIII | 66.81% | **19.54%** | 24.61% |
| | MCTS | 65.95% | 17.94% | **32.41%** |

Table 15. Comparison of search performance of various algorithms on the CIFAR_CNN in the CIFAR10 dataset

| Model/Dataset | Strategies | KMNC | NBC | SNAC |
|---|---|---|---|---|
| CIFAR-CNN CIFAR10 | Initial | 25.42% | 4.89% | 9.14% |
| | SPEA2 | **50.45%** | 10.39% | 15.27% |
| | NSGAIII | 43.61% | **10.86%** | **16.12%** |
| | MCTS | 47.52% | 8.83% | 15.25% |

Table 16. Comparison of search performance of various algorithms on the VGG16 in the ImageNet dataset

| Model/Dataset | Strategies | KMNC | NBC | SNAC |
|---|---|---|---|---|
| VGG16 ImageNet | Initial | 24.59% | 4.51% | 6.19% |
| | SPEA2 | **34.86%** | 8.39% | 13.07% |
| | NSGAIII | 32.27% | 9.14% | 12.71% |
| | MCTS | 34.12% | **9.58%** | **14.39%** |

Table 17. Comparison of search performance of various algorithms on the VGG19 in the ImageNet dataset

| Model/Dataset | Strategies | KMNC | NBC | SNAC |
|---|---|---|---|---|
| VGG19 ImageNet | Initial | 22.27% | 5.32% | 7.64% |
| | SPEA2 | 32.29% | **10.04%** | **11.01%** |
| | NSGAIII | **32.7%** | 8.71% | 9.6% |
| | MCTS | 31.73% | 9.53% | 9.87% |

Tables 12 to 17 show the coverage rate of various neuron coverage criteria across multiple neural networks and datasets using SPEA2, NSGAIII, and MCTS. The many-objective optimization algorithms show certain improvements over the MCTS method, achieving higher coverage for the three testing criteria in most cases. The MCTS-based DNN fuzz test generation method, which selects a mutation combination and then evaluates coverage rate, only continuing with new mutation combinations if there is an increase in coverage, is a local search method. Generally, it performs better than random search but is less effective than many-objective optimization algorithms.

Comparing the experimental results of SPEA2 and NSGAIII algorithms, SPEA2 appears more effective, surpassing NSGAIII on KMNC and SNAC criteria across multiple neural networks and datasets. For the KMNC criteria on LeNet4 and CIFAR_CNN networks, the SPEA2 algorithm achieves up to about 6% higher coverage rate compared to NSGAIII; similarly, for the SNAC criteria on LeNet4 and LeNet5 networks, it also increases by about 6%. For the NBC criteria, SPEA2 performs slightly worse than NSGAIII on few network models, but the maximum difference in coverage rate across all networks does not exceed 3%. For the VGG series networks, due to their complexity and fewer samples, the overall improvement in coverage rate is not significant, resulting in minor differences between strategies for various evaluation criteria.

### 4.2.3. Performance of Test Case Generation using improved SPEA2

To validate the effectiveness of the SPEA2 algorithm improved by the MCTS local search strategy and the decomposition-based archiving scheme, experiments were conducted comparing the DNN test generation framework with the improved SPEA2 algorithm, and with the framework with NGSAIII, as well as two popular neural network testing frameworks, DeepHunter and Tensorfuzz. Experimental results are shown in Tables 18 to 23.

Table 18. Coverage improvement comparison of various frameworks on the LeNet1 in the MNIST dataset

| Model/Dataset | Framework | KMNC | NBC | SNAC |
|---|---|---|---|---|
| LeNet-1 MNIST | Initial | 37.79% | 6.25% | 6.25 |
| | DeepSearch(MSPEA2) | 86.47% | **45.51%** | 45.08% |
| | DeepSearch(DSPEA2) | **87.42%** | 40.21% | 41.26% |
| | DeepSearch(NSGAIII) | 79.58% | 35.92% | 39.91% |
| | DeepSmartFuzz | 80.15% | 44.1% | **45.61%** |
| | DeepHunter | 84.04% | 42.78% | 42.37% |
| | TensorFuzz | 75.92% | 32.9% | 31.78% |

Table 19. Coverage improvement comparison of various frameworks on the LeNet4 in the MNIST dataset

| Model/Dataset | Framework | KMNC | NBC | SNAC |
|---|---|---|---|---|
| LeNet-4 MNIST | Initial | 39.57% | 9.51% | 14.08% |
| | DeepSearch(MSPEA2) | **89.18%** | **28.74%** | **47.83%** |
| | DeepSearch(DSPEA2) | 84.72% | 26.65% | 44.63% |
| | DeepSearch(NSGAIII) | 81.7% | 27.73% | 35.61% |
| | DeepSmartFuzz | 81.91% | 27.43% | 41.02% |
| | DeepHunter | 79.04% | 26.4% | 41.67% |
| | TensorFuzz | 70.99% | 19.71% | 33.82% |

Table 20. Coverage improvement comparison of various frameworks on the LeNet5 in the MNIST dataset

| Model/Dataset | Framework | KMNC | NBC | SNAC |
|---|---|---|---|---|
| LeNet-5 MNIST | Initial | 36.27% | 8.78% | 15.27% |
| | DeepSearch(MSPEA2) | **83.63%** | **31.47%** | **48.21%** |
| | DeepSearch(DSPEA2) | 79.43% | 30.19% | 43.36% |
| | DeepSearch(NSGAIII) | 79.5% | 29.1% | 42.11% |

| Model/Dataset | Framework | KMNC | NBC | SNAC |
|---|---|---|---|---|
| | DeepSmartFuzz | 77.94% | 26.06% | 44.31% |
| | DeepHunter | 71.74% | 22.62% | 42.16% |
| | TensorFuzz | 69.89% | 27.49% | 34.83% |

Table 21. Coverage improvement comparison of various frameworks on the CIFAR_CNN in the CIFAR10 dataset

| Model/Dataset | Framework | KMNC | NBC | SNAC |
|---|---|---|---|---|
| | Initial | 25.42% | 4.89% | 9.14% |
| | DeepSearch(MSPEA2) | **62.71%** | **37.41%** | **25.66%** |
| | DeepSearch(DSPEA2) | 57.04% | 28.36% | 23.22% |
| CIFAR-CNN CIFAR10 | DeepSearch(NSGAIII) | 54.38% | 29.87% | 22.67% |
| | DeepSmartFuzz | 56.95% | 33.72% | 21.96% |
| | DeepHunter | 53.35% | 29.87% | 21.43% |
| | TensorFuzz | 50.12% | 23.6% | 16.41% |

Table 22. Coverage improvement comparison of various frameworks on the VGG16 in the ImageNet dataset

| Model/Dataset | Framework | KMNC | NBC | SNAC |
|---|---|---|---|---|
| | Initial | 24.59% | 4.51% | 6.19% |
| | DeepSearch(MSPEA2) | **50%** | **17.44%** | 19.71% |
| | DeepSearch(DSPEA2) | 47.22% | 15.47% | 18.65% |
| VGG16 ImageNet | DeepSearch(NSGAIII) | 44.71% | 14.74% | 19.47% |
| | DeepSmartFuzz | 48.37% | 15.98% | **19.98%** |
| | DeepHunter | 41.06% | 12.33% | 16.76% |
| | TensorFuzz | 39.98% | 11.43% | 17.66% |

Table 23. Coverage improvement comparison of various frameworks on the VGG19 in the ImageNet dataset

| Model/Dataset | Framework | KMNC | NBC | SNAC |
|---|---|---|---|---|
| | Initial | 22.27% | 5.32% | 7.64% |
| | DeepSearch(MSPEA2) | **49.88%** | **20.18%** | **21.07%** |
| | DeepSearch(DSPEA2) | 45.91% | 17.76% | 17.16% |
| VGG19 ImageNet | DeepSearch(NSGAIII) | 47.11% | 15.58% | 17.21% |
| | DeepSmartFuzz | 45.65% | 17.9% | 19.41% |
| | DeepHunter | 42.91% | 16.27% | 14.87% |
| | TensorFuzz | 41.61% | 14.69% | 15.38% |

Tables 18 to 23 summarize the result of comparison in various coverage criteria for the fuzzing testing framework improved by multiple strategies and frameworks. From the experimental results, the following conclusions can be drawn:

(1) The fuzz testing framework using SPEA2 with MCTS local search strategy generally performs better. The MCTS strategy ensures the viability of individuals by making minor modifications to the chromosome, avoiding the problem of the population falling into local optima due to image semantic constraints. The SPEA2 framework improved by the decomposition-based archiving strategy can enhance the algorithm's global search capability, but overall, its effectiveness is still inferior to DeepSearch(MSPEA2). DeepSmartFuzz, mainly optimizing single criteria, performs better than many-objective optimization algorithms in some cases. DeepHunter and Tensorfuzz, although including new mutation strategies such as image noise, flips, and shifts, use random search strategies, resulting in higher efficiency but lower overall coverage compared to other frameworks.

(2) As network models and datasets become increasingly complex, the coverage rate in various testing criteria also gradually decreases. For the LeNet series networks, the KMNC criteria achieved up to about a 50% increase in coverage rate, while SNAC and NBC criteria also improved by around 40%. For the CIFAR_CNN network, the KMNC coverage rate increased by about 37%, while SNAC and NBC improved by about 16% and 33%, respectively. For the VGG series network models, due to fewer overall samples relative to the entire validation set, the coverage rate improvement is lower, with KMNC increasing by about 20% and SNAC and NBC by about 10%. As SNAC and NBC criteria mainly calculate the proportion of neurons whose output values are outside the main functional area, their coverage rate improvements are lower compared to the KMNC criteria

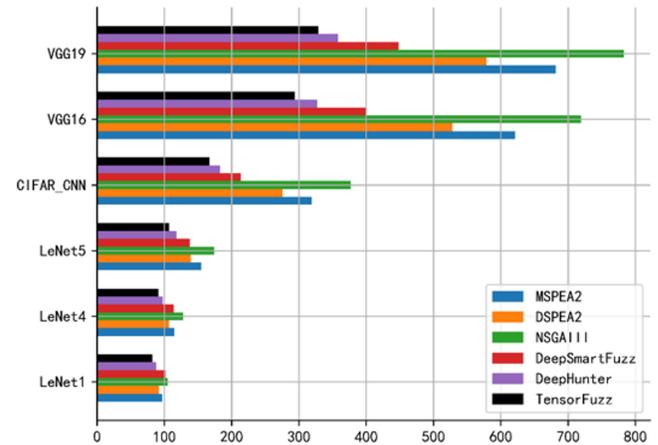

Figure 5. Time (m) consumption statistics for increased coverage rate for different methods on various networks

Figure 5 shows the time consumption statistics for each network model when calculating the coverage on different datasets. Since testing frameworks like DeepSmartFuzz optimize based on a single criterion, this paper selects the longest time consumed among all coverage criteria for their time statistics. By analyzing the experimental results, the following two conclusions can be drawn:

(1) As shown in Figure 5, the larger the network model and the more neurons it has, the greater the time consumption. Since the LeNet series network models use the same dataset and have similar structures, the overall time difference is not significant. The time difference becomes more apparent from the LeNet series to CIFAR_CNN, as the image channels increase from 1 to 3 and the network model becomes more complex. From CIFAR_CNN to the VGG series, with image sizes increasing from 32x32x3 to 244x244x3 and more model parameters and neurons, the overall time difference becomes even more pronounced.

(2) Comparing different frameworks, Tensorfuzz and DeepHunter have the shortest time consumption because they do not search the entire mutation space but directly perform random mutations, resulting in less improvement in coverage than other frameworks. The fuzz testing frameworks based on many-objective search and the DeepSmartFuzz framework have a small time difference in the LeNet series networks, not exceeding 20 minutes; however, the time difference is more significant in CIFAR_CNN and VGG series networks; in VGG19, the maximum difference approaches 250 minutes, with NSGAIII taking the longest, nearly 800 minutes. The SPEA2 framework with decomposition-based archiving strategy takes less time than the SPEA2 with MCTS local search, reducing about 100 minutes in VGG networks. It is evident that fuzz testing frameworks based on many-objective search are more effective in increasing neuron coverage but are less efficient. Further exploration and research into more efficient search algorithms are needed to enhance the efficiency of these frameworks

### 4.2.4. Correlation Between Generated Test and Adversarial Input Samples

To further verify the effectiveness of the test samples generated by the framework, this paper conducts further experiments on the generated test samples, counting the number and proportion of adversarial samples in the newly generated test inputs to verify its effectiveness. To ensure the fairness of the experiments, 16 samples were taken per experiment, selecting 64 images per sample. Specific experimental results are shown in Tables 24 to 29.

Table 24. Number of adversarial samples generated by different frameworks on the LeNet1

| Model/Dataset | Framework | KMNC | SNAC | NBC | Percent |
|---|---|---|---|---|---|
| LeNet-1 MNIST | DeepSearch | 87 | 87 | 87 | 8.4% |
| | DeepSmartFuzzer | 92 | 49 | 41 | 8.9% |
| | DeepHunter | **101** | 76 | 61 | **9.8%** |
| | TensorFuzz | 78 | 24 | 11 | 7.6% |

Table 25. Number of adversarial samples generated by different frameworks on the LeNet4

| Model/Dataset | Framework | KMNC | SNAC | NBC | Percent |
|---|---|---|---|---|---|
| LeNet-4 MNIST | DeepSearch | **74** | **74** | **74** | **7.2%** |
| | DeepSmartFuzzer | 72 | 26 | 34 | 7% |
| | DeepHunter | 51 | 42 | 54 | 5.2% |
| | TensorFuzz | 30 | 24 | 28 | 2.9% |

Table 26. Number of adversarial samples generated by different frameworks on the LeNet5

| Model/Dataset | Framework | KMNC | SNAC | NBC | Percent |
|---|---|---|---|---|---|
| LeNet-5 | DeepSearch | 62 | 62 | 62 | 6% |

| MNIST | DeepSmartFuzzer | 48 | 26 | 45 | 4.6% |
|---|---|---|---|---|---|
| | DeepHunter | 43 | 31 | 37 | 4.2% |
| | TensorFuzz | 21 | 9 | 13 | 2% |

Table 27. Number of adversarial samples generated by different frameworks on the CIFAR_CNN

| Model/Dataset | Framework | KMNC | SNAC | NBC | Percent |
|---|---|---|---|---|---|
| CIFAR-CNN CIFAR10 | DeepSearch | 662 | 662 | 662 | 64.6% |
| | DeepSmartFuzzer | 610 | **676** | 621 | **66%** |
| | DeepHunter | 622 | 629 | 617 | 61.4% |
| | TensorFuzz | 601 | 614 | 622 | 60.7% |

Table 28. Number of adversarial samples generated by different frameworks on the VGG16

| Model/Dataset | Framework | KMNC | SNAC | NBC | Percent |
|---|---|---|---|---|---|
| VGG16 imageNet | DeepSearch | 547 | 547 | 547 | **53.41%** |
| | DeepSmartFuzzer | 531 | 493 | 501 | 51.85% |
| | DeepHunter | 507 | 457 | 481 | 49.51% |
| | TensorFuzz | 481 | 468 | 472 | 46.97% |

Table 29. Number of adversarial samples generated by different frameworks on the VGG19

| Model/Dataset | Framework | KMNC | SNAC | NBC | Percent |
|---|---|---|---|---|---|
| VGG19 imageNet | DeepSearch | 535 | 535 | 535 | **52.24%** |
| | DeepSmartFuzzer | 517 | 498 | 509 | 50.48% |
| | DeepHunter | 472 | 505 | 483 | 49.31% |
| | TensorFuzz | 484 | 479 | 496 | 48.43% |

From Tables 24 to 29, it is evident that the number and proportion of adversarial samples generated by various neural network models in different frameworks on their respective datasets are influenced by the complexity of the dataset and the coverage criteria used, as well as the specific neural network model. Three key conclusions can be drawn from the experimental results:

(1) For the LeNet series of network models, the number of adversarial samples generated is relatively small, with the proportion not exceeding 10%, and in some cases, as low as 2%. This is primarily because the LeNet series has fewer layers and parameters, and achieves a very high level of accuracy. Additionally, the MNIST dataset used with these models is relatively simple, limiting the generation of a larger number of adversarial samples. In contrast, the 20-layer CIFAR_CNN network model, which is more complex and uses the larger CIFAR10 dataset, generates a greater number of adversarial samples. For the VGG series of networks, fewer samples were extracted, and as previous experiments have shown, the coverage improvement is less than other network models, resulting in a lower number of adversarial samples generated compared to the CIFAR_CNN.

(2) Across multiple coverage criteria, it is observed that the k-multisection neuron coverage (KMNC) generates more adversarial samples for most models than the other two coverage criteria (NBC and SNAC). The KMNC tests the main functional area of neurons, making it more effective both in terms of coverage improvement and in the generation of adversarial samples. However, in some cases, the NBC and SNAC criteria produced better results, as shown in Tables 27, 28, and 29.

(3) Comparing various experimental frameworks and testing methods, the SPEA2-based DNN test generation, improved through multiple proposed strategies, generally outperforms the other frameworks. The Tensorfuzz framework, which primarily uses data noise for image mutation, is less effective than the brightness, contrast, and blurring mutation strategies used in this study. The DeepHunter framework, while introducing additional mutation strategies such as image translation and rotation, can sometimes violate the semantic constraints of the images, resulting in better performance in some scenarios, as shown in Table 24. The correlation between high coverage rates and the number of adversarial samples generated suggests that the methods and improvement strategies proposed in this study not only enhance neuron coverage rates but also generate adversarial samples that can expose errors in neural network models.

### 4.2.5. Threats to Validity

The algorithm's configurations and hyperparameter's settings in the coverage criteria proposed in this paper may affect the validity of the results. Although multiple coverage criteria have been considered, not all configurable parameters have been covered, such as the k-value of the k-multisection neuron coverage, which may not represent the best parameter selection. To mitigate these threats, multiple repetitions of each experiment set were conducted to avoid the impact of experimental randomness. Regarding algorithm parameter settings, the framework used in this paper experimented with multiple parameters to select the best configuration, while comparison frameworks used the same parameters as in their respective papers.

Also, the choice of evaluation objects (i.e., DNN network models and datasets) might pose a threat to the universality of the experimental results. In terms of datasets, this paper chose the commonly studied MNIST handwritten dataset, CIFAR10 dataset, and the large-scale ImageNet dataset. For each dataset, popular network models were selected, including the LeNet series and the more complex VGG series models, with neuron counts ranging from dozens to tens of thousands. Nevertheless, the results of this research might not be generalizable to other datasets and network models, and more comprehensive evaluations may needed in future work.

### 5. CONCLUSION

This paper first proposes a frequency-based fuzz data sampling strategy for DNN test generation. The proposed strategy has been validated to outperform random sampling in most cases and is generally better than cluster-based sampling strategies. Next, a SPEA2-based method for DNN test case generation is proposed, optimizing multiple states of neuron activation values at same time. To address its deficiency in local search, a Monte Carlo Tree Search-based local search algorithm is designed to generate test cases with higher coverage. Furthermore, the paper uses a decomposition-based archive strategy to improve SPEA2-based DNN test generation, dividing the solution space with different weight vectors and establishing corresponding relationships, enhancing global search capabilities. By dynamically adjusting the size of the algorithm's external archive, the computational efficiency is improved. Testing on multiple datasets and network models has proven the superiority of proposed method for DNN test case generation. Finally, the paper conducts a correlation study between test cases generated by the proposed method and adversarial samples. It shows that generated test cases with high coverage can effectively discover defects in neural networks and generate adversarial samples that cause errors in the networks. Future studies of the paper may include, but are not limited to: 1) Extensibility of neural networks, extending DNN test generation for speech recognition, audio and natural language, through different mutation strategies. 2) Sufficiency of neural network testing, considering only neuron coverage is somewhat limited; isolated neuron coverage may not cover all the logic of a neural network model, so exploring new coverage criteria is necessary. 3) Efficiency of neural network testing, as models and datasets become more complex, computational time increases. Improving the efficiency of DNN test generation is important, e.g., prioritizing test inputs.